# High-Fidelity Modelling of the Molten Salt Fast Reactor

**Maximiliano Dalinger, Elia Merzari, Saya Lee, and Casey Emler**
The Pennsylvania State University
mgd5394@psu.edu; ebm5351@psu.edu; sayalee@psu.edu; cae5292@psu.edu

**ABSTRACT**

The Molten Salt Fast Reactor (MSFR) is one of the six GEN-IV reactor designs. In the MSFR, the liquid fuel is the coolant, which moves throughout the primary circuit. This complex phenomenology requires multiphysics modeling. In the present paper, a model of the MSFR is developed in the multiphysics code Cardinal, considering neutronic-thermal hydraulic feedback and the transport of delayed neutron precursors (DNPs) and decay heat precursors (DHPs). OpenMC is used to solve neutronic equations, and NekRS is used to solve mass, momentum, energy, DNPs, and DHPs distribution. A RANS $k - \tau$ turbulence model is used in NekRS. DNPs and DHPs are modeled using a convective-diffusion equation with modified source terms considering radioactive decay. Cardinal results showed a reasonable behavior for temperature, heat source, velocity, DNPs, and DHPs. However, the current limitations in OpenMC do not allow the modification of delayed neutron source locations. Ongoing efforts look to include this feature in future work to introduce DNP feedback in OpenMC.

## 1. INTRODUCTION

The Molten Salt Reactor (MSR) was selected in 2002 at The Generation-IV International Forum as one of the reference reactor concepts because of its enhanced safety and reliability, reduced waste generation, effective fuel use, and improved economic competitiveness [1]. Since then, an innovative concept called the Molten Salt Fast Reactor (MSFR) has been proposed. This design eliminates the graphite moderator, resulting in a fast breeder reactor. The MSFR design has the particularity that the fuel is the coolant itself, which produces a tight coupling between neutronics and thermal hydraulics as the fuel circulates through the primary system. Therefore, developing computational models for the analysis of the MSFR requires a multi-physics approach. The fission process generates fission products, some of which decay, releasing decay heat and delayed neutrons, which is why they are called delayed neutron precursors (DNPs) and decay heat precursors (DHPs), respectively. In the MSFR, these precursors originate and are carried by the liquid fuel throughout the primary circuit. The generation, transport, and decay of the DNPs affect the neutron flux, heat source, and temperature distributions in the MSFR.

Previous works modeled the MSFR in Cardinal but without considering the delayed neutron precursors drift [2] or considering a deterministic approach to solve the neutron transport equation [3]. In the present paper, we propose to develop a neutronic–thermal hydraulics computational model of the MSFR that considers the transport of the delayed neutron and decay heat precursors along the primary circuit. The principal computational tool chosen for this purpose is the high-fidelity code Cardinal, a wrapping within the MOOSE framework that integrates the Computational Fluid Dynamics code NekRS and the Monte Carlo particle transport code OpenMC.

## 2. THE MOLTEN SALT FAST REACTOR

The Safety Assessment of the Molten Salt Fast Reactor (SAMOFAR) project was created to advance experimental and numerical techniques relating to Molten Salt Reactor technologies. The Molten Salt Fast Reactor (MSFR) design was developed. The MSFR is a 3000 MWth liquid fuel reactor with a cylindrical core of 2.25 m high and a diameter of 2.25 m. The reactor system is divided into three distinct circuits: the primary fuel circuit, the intermediate circuit, and the power conversion circuit. The primary fuel circuit contains 18 m$^3$ of a molten fuel salt LiF-ThF$_4$-$^{233}$UF$_4$ blend with 77.5 mol% of LiF [5]. Figure 1 shows the MSFR core layout. The reactor has 16 sets of inlet and outlet pipes equally spaced around the core, each equipped with pumps and heat exchangers. Half of the total salt volume is inside the core at any time. The fuel salt flows from the bottom to the top of the core and is fed into the heat exchangers and pumps. A stainless-steel reflector surrounds the reactor. A breeder blanket of LiF-$^{232}$ThF$_4$ with 77.5 mol% of LiF surrounds the core radially. A layer of B$_4$C surrounds the fertile blanket to absorb any remaining neutrons and protect the pumps and heat exchangers.

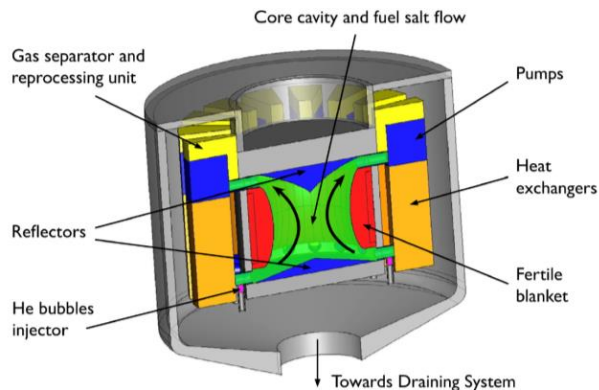

**Figure 1: Design of SAMOFAR MSFR.**

## 3. COMPUTATIONAL TOOLS

### 3.1. OpenMC

OpenMC [6] is an open-source high-fidelity Monte Carlo neutron and photon transport simulation code. It can perform k-eigenvalue and fixed source calculations on models built using constructive solid geometry (CSG) or computer aid design (CAD) representation. Some of its features include continuous-energy transport, Doppler broadening of cross-sections, and calculation of reaction and recoverable energy production rates due to fission.

### 3.2. NekRS

NekRS [7] is a GPU-accelerated open-source Spectral Element Method (SEM) Computational Fluid Dynamics (CFD) code to simulate fluid dynamics and heat transfer. It solves the unsteady incompressible Navier-Stokes and energy equations in 2-D or 3-D geometries. Its capabilities include solving Conjugated Heat Transfer (CHT) problems, Reynolds-Averaged Navier Stokes (RANS) modeling, Large Eddy Simulation (LES), and Direct Numerical Simulation (DNS). It supports both CPU and GPU backends. In this work, we use the RANS $k - \tau$ turbulence model. NekRS can also model the transport of passive scalars using a convection-diffusion equation. This model can be used when changes in scalar concentrations do not influence the physical properties of the flow.

### 3.3. MOOSE

MOOSE [8] (Multiphysics Object-Oriented Simulation Environment) is a finite element framework for solving fully coupled, fully implicit multiphysics simulations. It executes multiple sub-applications simultaneously and transfers data between the scales. It discretizes the space using the PETSc non-linear solver and libMesh. Some of its capabilities include automatic differentiation, scaling to a large number of processors, hybrid parallelism, and mesh adaptivity.

### 3.4. Cardinal

Cardinal [9] is an open-source application that wraps OpenMC and NekRS codes within the MOOSE framework. It uses the MOOSE data transfer implementation to perform high-fidelity coupled neutronic-thermal hydraulics calculations. It can also be coupled with any MOOSE-based application, enabling broad multiphysics capabilities.

### 4. FORMULATIONS

OpenMC solves the neutron transport equation by performing a k-eigenvalue calculation. It calculates the recoverable energy from fission $q'''_{fis}$ using a *kappa-fission* tally and the total production of neutrons due to fission $R_{vf}$ using a *nu-fission* tally. These quantities are calculated by OpenMC as shown in Equations 1 and 2, where $E_{f,g}$ is the average energy released by fission, $\Sigma_{f,g}$ is the fission cross-section, $\phi_g$ is the neutron flux, and $v\Sigma_{f,g}$ is the nu-fission cross-section of energy group $g$.

$$q'''_{fis} = \sum_g E_{f,g} \Sigma_{f,g} \phi_g \qquad (1)$$

$$R_{vf} = \sum_g v\Sigma_{f,g} \phi_g \qquad (2)$$

NekRS solves fluid mass, momentum, and energy conservation equations with an incompressible RANS model, shown in Equations 3, 4, and 5, respectively, where $\vec{u}$ is the velocity, $\rho$ is the density, $p$ is the pressure, $\mu_l$ is the laminar dynamic viscosity, $\mu_t$ is the turbulent dynamic viscosity, $Cp$ is the specific heat capacity, $T$ is the temperature, $k_l$ is the laminar thermal conductivity, $k_t$ is the turbulent thermal conductivity, and $q'''$ is the volumetric heat source. The turbulent Prandtl number $Pr_t$ relates $\mu_t$ and $k_t$, as shown in Equation 6.

$$\nabla \cdot \vec{u} = 0 \tag{3}$$

$$\rho \left(\frac{\partial \vec{u}}{\partial t} + \vec{u} \cdot \nabla \vec{u}\right) = -\nabla p + \nabla \cdot \left((\mu_l + \mu_t) \nabla \vec{u}\right) \tag{4}$$

$$\rho\, Cp \left(\frac{\partial T}{\partial t} + \vec{u} \cdot \nabla T\right) = \nabla \cdot \left((k_l + k_t) \nabla T\right) + q''' \tag{5}$$

$$Pr_t = \frac{\mu_t\, Cp}{k_t} \tag{6}$$

The $k - \tau$ turbulence RANS model evaluates $\mu_t$ based on two additional partial differential equations for the turbulent kinetic energy $k$ and the inverse specific dissipation $\tau$. NekRS's RANS model is currently limited to constant fluid properties, so $\rho$, $\mu_l$, $Cp$, and $k_l$ are assumed constant.

The circulating liquid fuel in the MSFR allows delayed neutron precursors to move. Therefore, transport equations are formulated also for Delayed Neutron Precursors (DNPs) and Decay Heat Precursors (DHPs) [10]. DNPs and DHPs are treated as passive scalars in NekRS using a modified convection-diffusion equation, incorporating decay and source terms. The DNP concentration $c_i$ of group $i$ is calculated as shown in Equation 7, where $\lambda_{d,i}$ is the decay constant, and $\beta_{d,i}$ is the delayed neutron fraction of DNP group $i$. An analogous equation is used for the DHP concentration $d_l$ of the group $l$ as shown in Equation 8, where $\lambda_{h,l}$ is the decay constant and $\beta_{h,l}$ is the decay heat fraction of the DHP group $l$. Here, $d_l$ represents the volumetric "latent" fission energy. Therefore, the volumetric heat source in Equation 5 can be defined as shown in Equation 9.

$$\frac{\partial c_i}{\partial t} + \nabla \cdot (\vec{u}\, c_i) = \nabla \cdot \left(D_{eff}\, \nabla c_i\right) - \lambda_{d,i}\, c_i + \beta_{d,i}\, R_{vf} \tag{7}$$

$$\frac{\partial d_l}{\partial t} + \nabla \cdot (\vec{u}\, d_l) = \nabla \cdot \left(D_{eff}\, \nabla d_l\right) - \lambda_{h,l}\, d_l + \beta_{h,i}\, q'''_{fis} \tag{8}$$

$$q''' = (1 - \beta_h)\, q'''_{fis} + \sum_l \lambda_{h,l}\, d_l \tag{9}$$

The diffusion coefficient $D_{eff}$ is defined as shown in Equation 10, where $Sc$ is the Schmidt number, $Sc_t$ is the turbulent Schmidt number, $\nu$ the kinematic viscosity, and $\nu_t$ is the turbulent kinematic viscosity.

$$D_{eff} = \frac{\nu}{Sc} + \frac{\nu_t}{Sc_t} \tag{10}$$

## 5.  COMPUTATIONAL MODEL

Calculations used simplified geometry. Figure 1 shows that the MSFR design has pumps and heat exchangers; however, these are not included in the model for simplification. Parameters used in coupled calculations are summarized in Table I. Fluid properties are calculated at an average temperature $T_{avg}$ of 973 K [5].

Table I. Parameters used in coupled calculations.

| Parameter | Symbol | Value |
|---|---|---|
| Fluid density [kg/m3] | $\rho$ | 4125.3 |
| Fluid dynamic viscosity [Pa.s] | $\mu_l$ | 0.01014673 |
| Fluid heat capacity [J/kg.K] | $Cp$ | 1593.9 |
| Fluid thermal conductivity [W/m.K] | $k_l$ | 1.009732 |
| Power [W] | $P$ | $3.0*10^8$ |
| Inlet Temperature [K] | $T_{in}$ | 898.0 |
| Outlet Temperature [K] | $T_{out}$ | 998.0 |
| Delta Temperature [K] | $\Delta T$ | 100.0 |
| Average Temperature [K] | $T_{avg}$ | 973.0 |
| Mass flow rate [kg/s] | $\dot{m}$ | 1882.12 |
| Inlet velocity [m/s] | $v_{in}$ | 0.1886 |
| Hydraulic diameter [m] | $D_h$ | 0.4 |

### 5.1. OpenMC MSFR Model

The OpenMC model was generated using a simplified cylindrical geometry using CSG techniques. The liquid fuel is LiF-ThF$_4$-$^{233}$UF$_4$, and the breeder blanket contains LiF-$^{232}$ThF$_4$. The blanket is surrounded by a boron carbide (B$_4$C) layer to absorb any remaining neutrons. Radial and axial reflectors surround the whole core, constructed of HT9 stainless-steel. Figure 2 shows the OpenMC geometry colored by material (A and B) and cell (C and D), where red represents the liquid fuel, blue is the fertile blanket, green is the absorber, and grey is the stainless-steel reflector. The fuel region is discretized to account for the change in fuel temperature and density. Temperature and density feedback is applied to the fuel region exclusively, while other components are always at 973K. The entire geometry is surrounded by a void cell with vacuum boundary conditions. Simulations used 1000000 particles, with 20 inactive batches and 100 total batches, obtained from a Shannon Entropy analysis, which resulted in uncertainties below 12.5 pcm for the effective multiplication factor *Keff*. The cross-section library used was ENDF/B-VIII.0. Current limitations of OpenMC do not allow us to modify the intensity and location where delayed neutrons are produced. There are ongoing efforts to be able to do so using DNP distributions.

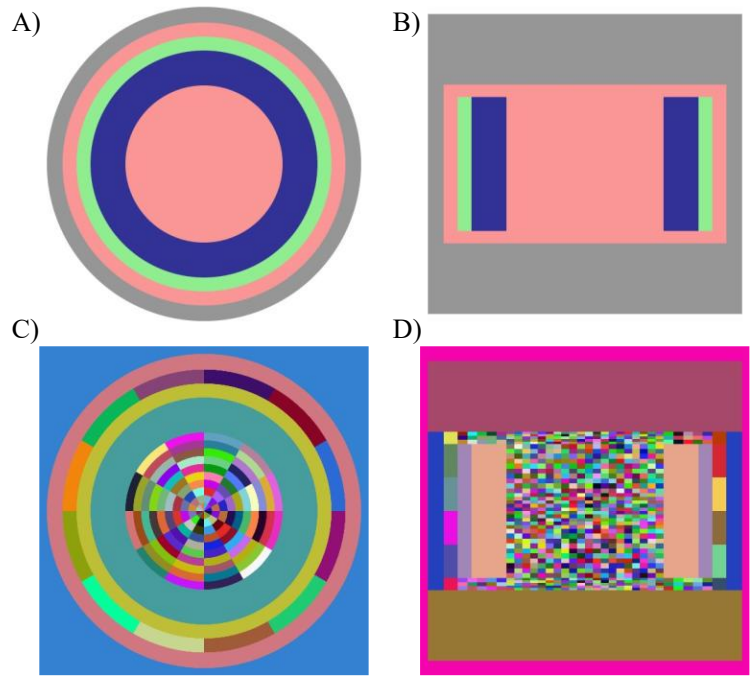

**Figure 2: OpenMC model of the MSFR. A) upper view colored by material, B) lateral view colored by material, C) upper view colored by cell, and D) lateral view colored by cell.**

### 5.1.1. Meshing analysis in OpenMC

A meshing analysis was performed in OpenMC using a uniform fuel temperature and density. Figure 3 shows the results of the meshing analysis in the axial direction for the edge, middle, and central regions of the central cylinder, as well as the radial meshing of the central cylinder. Each point corresponds to the average heat source over each cell for the centerline in the axial case and a radial line in the radial case. In the figure, the orange square represents the region where the meshing analysis is performed on each case. The analysis determined a discretization of 25 axial, 10 radial, and 12 azimuthal cells in the central cylinder; 3 axial, 5 radial, and 12 azimuthal cells for the superior and inferior region of the feeder; 3 axial, 1 radial, and 12 azimuthal cells for the vertical region of the feeder; resulting in a total of 5238 cells for the model. This analysis will be repeated in the future using non-uniform temperature and density profiles since, in MSFR, these profiles are not uniform, as seen below. An additional extensive meshing analysis of the feeders may be necessary if DNPs are included in OpenMC.

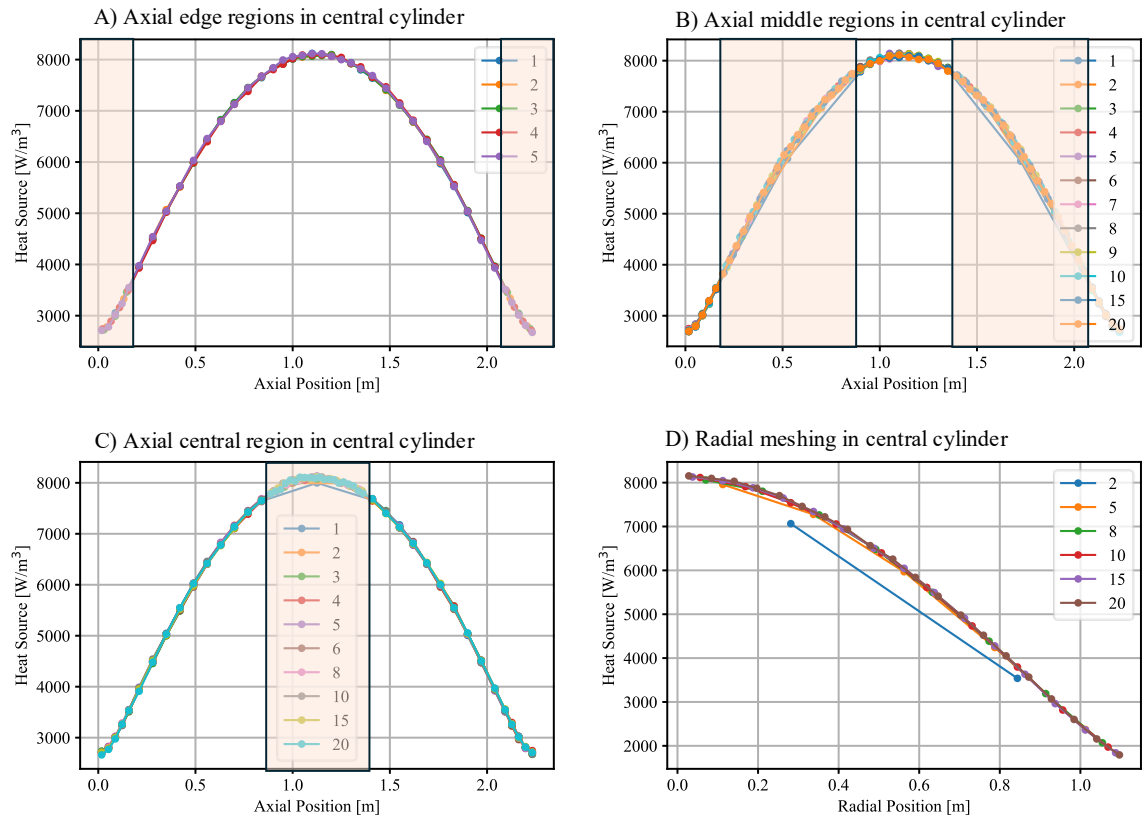

**Figure 3**: OpenMC meshing analysis. Orange blocks show the axial region selected for the analysis on each case. A) Axial edge regions in central cylinder, B) axial middle regions in central cylinder, C) axial central regions in central cylinder, and D) radial meshing in central cylinder.

### 5.1.2. MOOSE mesh to receive OpenMC results

When OpenMC results are transferred to Cardinal, these require a mesh that MOOSE can interpret, as shown in Figure 4. This geometry resembles the NekRS geometry but with the OpenMC meshing. It has been generated using MOOSE's Reactor Module and has 43200 cells. This geometry has been generated to match the original OpenMC mesh radially and axially. This mesh is also used to calculate each cell's average temperature and density values before transferring to OpenMC.

### 5.2. NekRS MSFR Model

NekRS solves fluid mass, momentum, and energy conservation using a RANS $k - \tau$ model for turbulence. It receives the volumetric heat source calculated in OpenMC. Figure 5 presents the three-dimensional geometry of the MSFR primary loop used for calculations, which was generated using the software Gmsh. The model uses simplified geometry with a continuous inlet/outlet and does not model the pumps or heat exchangers. The separation between the inlet and the outlet is 50 cm. Boundary conditions for the energy equation are uniform inlet temperature of 898K, adiabatic for walls, and the code calculates the outlet temperature. Regarding the momentum equation, boundary conditions are a parabolic velocity profile with a mean non-dimensional value of 1 for the inlet, non-slip for the remaining walls, and the code calculates the outlet velocity. For turbulence variables $k$ and $\tau$, parabolic profiles were used at the inlet with mean non-dimensional values of 0.01 and 0.1, respectively. Table II shows the relevant non-dimensional parameters used for simulations. The mesh has 1,356,800 hexahedral elements. A polynomial order 3 was selected, resulting in about $8.68*10^7$ Gauss-Lobatto-Legendre (GLL) quadrature point points. This resolution was found adequate for the problem.

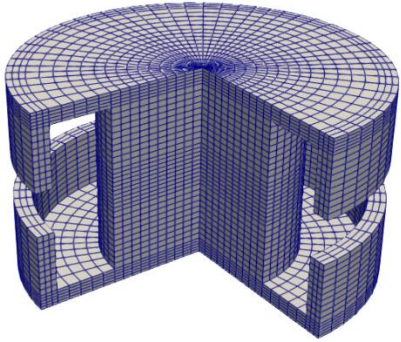

Figure 4: MOOSE mesh of the MSFR to receive OpenMC results.

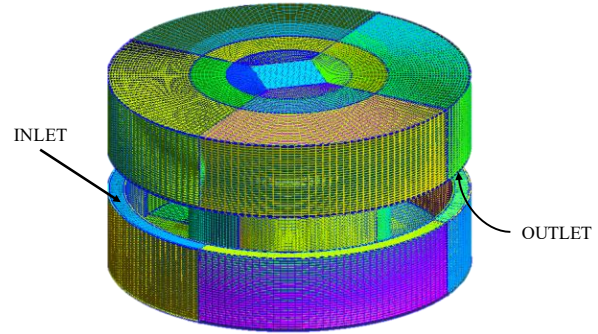

Figure 5: NekRS mesh of the MSFR.

DNPs and DHPs are modeled as passive scalars in NekRS. Table III summarizes constants used for 6 groups of DNPs [11] and 3 groups of DHPs [3]. To simulate the recirculation of DNPs and DHPs, when they reach the outlet, an average concentration is calculated and imposed at the inlet by a decay constant. This calculation assumes a residency time of 2 seconds in the pumps and heat exchangers. The *nu-fission* volumetric source used for DNPs is calculated in OpenMC and transferred to NekRS through Cardinal.

Table II. Nondimensional parameters used for NekRS coupled calculations.

| Nondimensional Parameter | Symbol | Value |
|---|---|---|
| Reynolds Number | $Re$ | 30672 |
| Prandtl Number | $Pr$ | 16.02 |
| Peclet Number | $Pe$ | 491286 |
| Turbulent Prandtl Number | $Pr_t$ | 1.00 |
| Schmidt Number | $Sc$ | 1.00 |
| Turbulent Schmidt Number | $Sc_t$ | 0.85 |

Table III. Constans used for DNPs [11] and DHPs [3].

| DNP Group | Decay Constant $\lambda_{n,i}$ [1/s] | Delayed Neutron Fraction $\beta_{n,i}$ [-] |
|---|---|---|
| 1 | 0.012575 | 0.0002666 |
| 2 | 0.033414 | 0.0008494 |
| 3 | 0.130755 | 0.0007037 |
| 4 | 0.302620 | 0.0009827 |
| 5 | 1.269231 | 0.0002263 |
| 6 | 3.135747 | 0.0000713 |
| DHP Group | Decay Constant $\lambda_{h,l}$ [1/s] | Decay Heat Fraction $\beta_{h,l}$ [-] |
| 1 | 0.197400 | 0.0117 |
| 2 | 0.016800 | 0.0129 |
| 3 | 0.000358 | 0.0186 |

### 5.3. Cardinal Multiphysics Coupling

Figure 6 presents the coupling procedure used in Cardinal, often called Picard iteration "in time" [9]. OpenMC runs a k-eigenvalue calculation using an initial uniform temperature and density. It calculates the heat source $q'''_{fis}$ and nu-fission reaction rate $R_{vf}$. Cardinal receives this information and sends it to NekRS. OpenMC calculates $q'''_{fis}$ and $R_{vf}$ in [eV/source particle] and [particles/source particle], and Cardinal converts them to [W/m³] and [1/m³s], respectively. Since Cardinal can only transfer a volumetric heat source for the energy equation, its source code was modified to transfer an additional volumetric source named volumetric *scalar source* to be able to use it in passive scalars. Then NekRS performs $N$ calculations of time step $\Delta t_{nek}$ to calculate fluid temperature, velocity, DNPs, and DHPs concentration. NekRS sends the fluid temperature and DNP concentrations to Cardinal. Cardinal calculates the fuel density [5] and sends both fuel temperature and density. Temperatures and numerical densities are updated in OpenMC, and the process is repeated until convergence. The convergence is reached when the maximum, the average, and the outlet temperature reach a steady state value. For the present work, $N = 4000$ and $\Delta t_{nek} = 2.0 * 10^{-4}$. Therefore, the OpenMC model receives feedback for the fuel temperature and density. Current limitations in OpenMC do not allow the modification of delayed neutrons source locations or intensity. Ongoing efforts look to include this feature in future work.

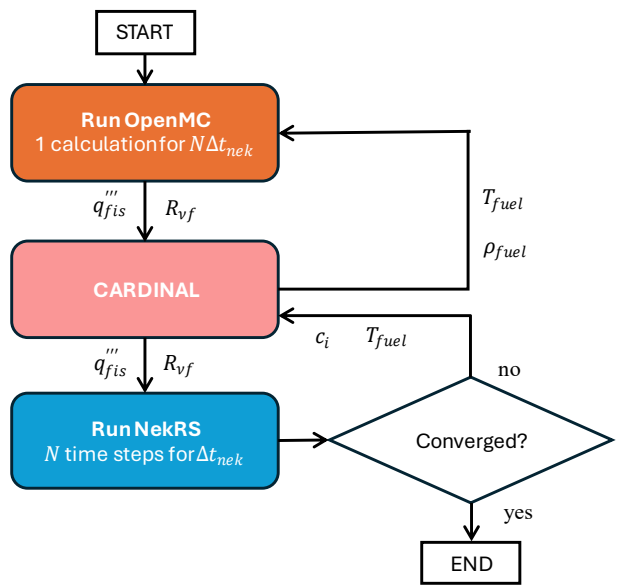

**Figure 6: Cardinal coupling procedure.**

### 6. SINGLE PHYSICS MODELS VERIFICATION

Single physics models have been verified against other codes. For OpenMC, results have been verified against a model developed in the Monte Carlo code Serpent 2 using a uniform temperature profile. A k-eigenvalue calculation was performed with 50000 particles, 20 inactive batches, and 100 total batches. Only one cell was used for the fuel in these models. Figure 7 compares results for the neutron flux in axial and radial directions. The multiplication factor $k_{eff}$ obtained with OpenMC was 1.04364 ± 0.00039, while in Serpent, it was 1.04338 ±0.00075. It is considered that there is a good agreement between codes.

A $k-\tau$ RANS model developed in Nek5000, a CPU version of NekRS, was verified against OpenFOAM CFD code using different turbulence models. The 2D geometry and parameters are slightly different from the 3D geometry used in the full coupling model to match reference [12]. The Reynolds number used is approximately 30000. Figure 8 compares the average velocity field obtained in both cases. Figure 9 presents the velocity magnitude along line A. There is good agreement between codes, as shown in both figures. The highest differences are observed around the peak centerline velocity, where Nek5000 velocities are higher than OpenFOAM. These discrepancies can potentially be explained by the different turbulence models used and variations in wall treatments between codes.

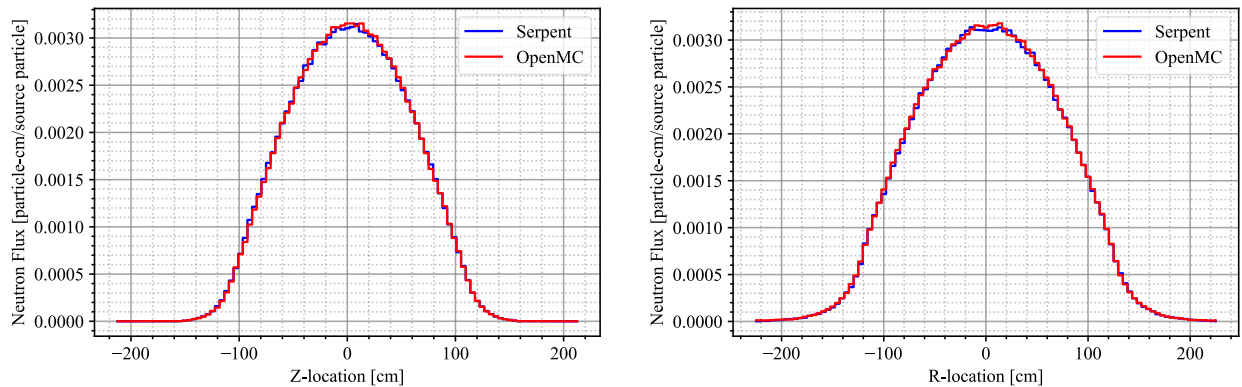

**Figure 7: Neutron Flux comparison between OpenMC and Serpent for axial position (left) and radial position (right)**

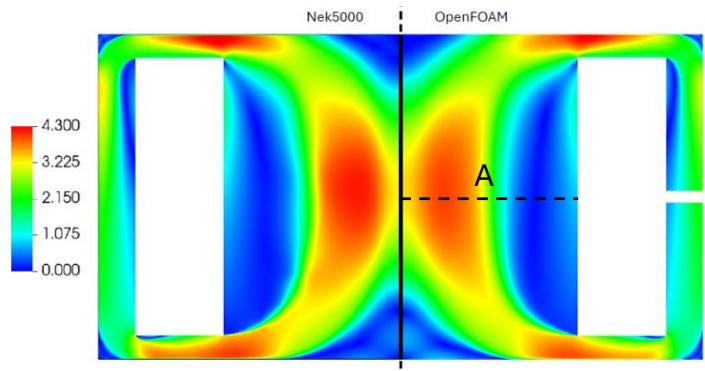

**Figure 8: Velocity field comparison between Nek5000 $k-\tau$ model and OpenFOAM $k-\omega$ model for a 2D geometry.**

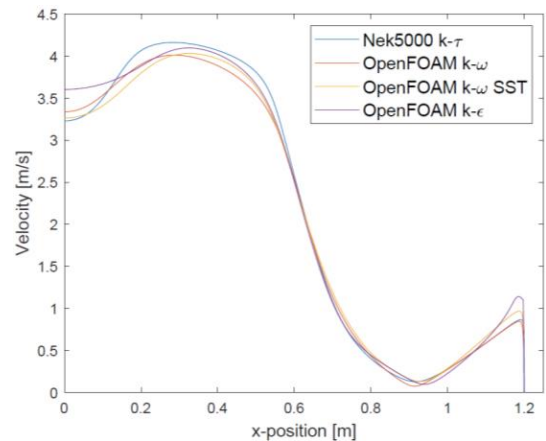

**Figure 9: Velocity distribution along line A between Nek5000 and OpenFOAM using different turbulence models.**

## 7. MULTIPHYSICS RESULTS

Simulations in Cardinal were performed on the supercomputer Frontier using GPUs for NekRS and CPUs for OpenMC. The results presented here are preliminary. The multiplication factor obtained is 1.068301 ± 0.00010. The simulation was performed for approximately 1000 convective units, and then a time-averaging was performed for approximately 100 more convective units. Figure 10 presents the results for the fission heat source and Figure 11 for the nu-fission reaction rate, both obtained in OpenMC. We see that both profiles have sinusoidal shapes. Figure 12 shows the time-averaged velocity distribution obtained in NekRS. We see stagnant regions at the center top and bottom and close to the lateral walls in the central cylinder. Figure 13 shows the time-averaged temperature distribution obtained in NekRS. We can see that the temperature rises in the stagnation regions of the flow. A rise in the fuel temperature is produced there because the flow is "trapped" and still being heated. The outlet temperature is 997.3K, 0.7K below the reference value. The average temperature is 983.1K, 10.1K higher than the reference value.

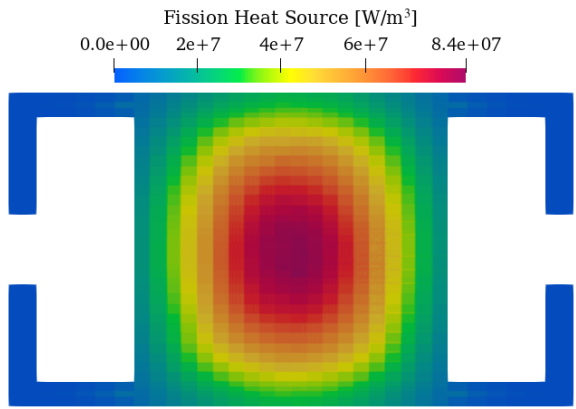

**Figure 10: Fission heat source calculated with OpenMC.**

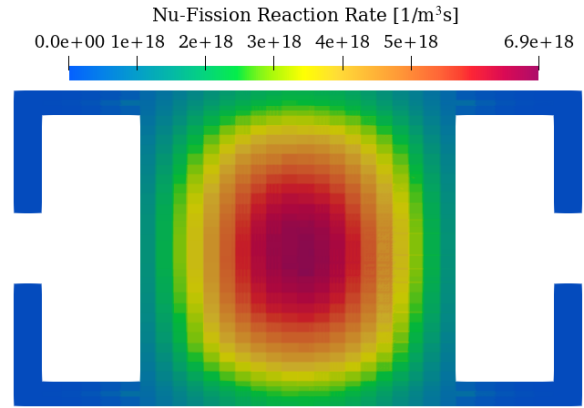

**Figure 11: Nu-fission reaction rate calculated with OpenMC.**

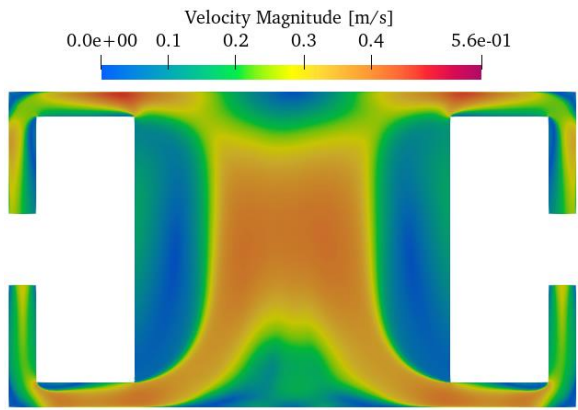

**Figure 12: Time-averaged velocity distribution calculated with NekRS.**

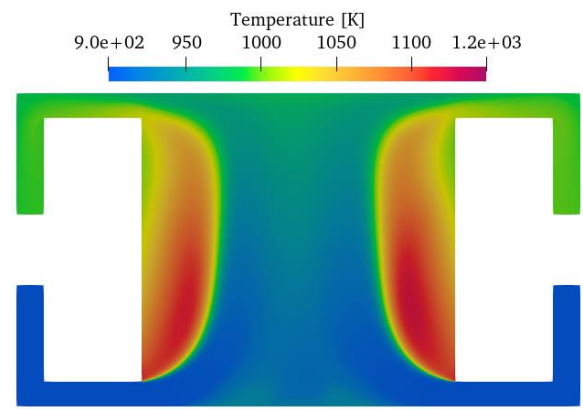

**Figure 13: Time-averaged temperature distribution calculated with NekRS.**

Time-averaged DNP concentrations for groups 1, 3, 5, and 6 are presented in Figures 14 to 17. For group 6, as shown in Figure 17, DNP distribution is like the nu-fission reaction rate. This distribution is because group 6 has the highest $\lambda_{n,i}$, meaning DNPs decay faster, close to the position where they were generated. As the $\lambda_{n,i}$ decreases, DNPs' decay time increases. This behavior is opposite to those seen for group 1 as shown in Figure 14, where DNPs move throughout the primary circuit, re-entering the core. Similar behavior is seen with time-averaged DHPs in Figures 18 and 19 for groups 1 and 3, respectively, but in this case, group 1 is the fastest in decay and group 3 the slowest. These results are preliminary and will be improved for the final version.

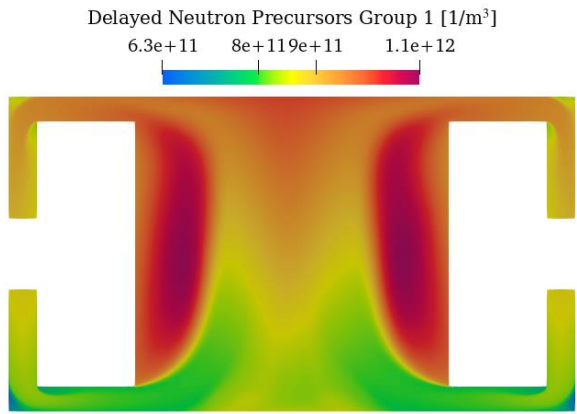

Figure 14: Time-averaged DNP concentration of group 1 calculated with NekRS.

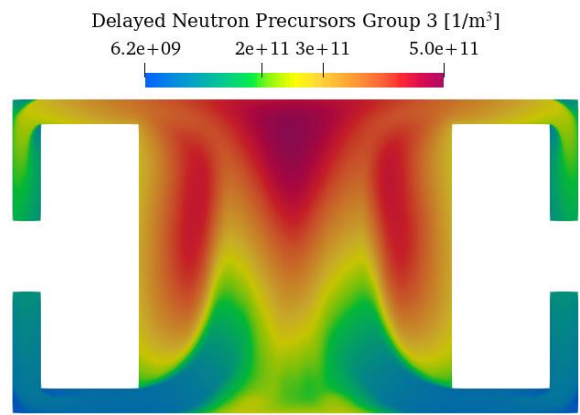

Figure 15: Time-averaged DNP concentration of group 3 calculated with NekRS.

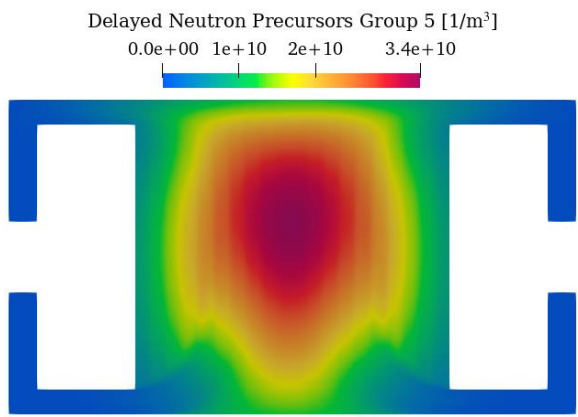

Figure 16: Time-averaged DNP concentration of group 5 calculated with NekRS.

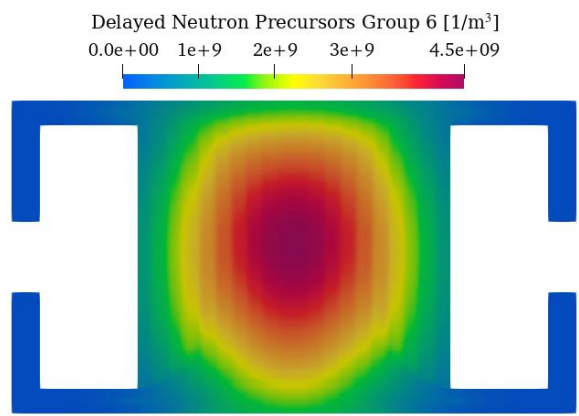

Figure 17: Time-averaged DNP concentration of group 6 calculated with NekRS.

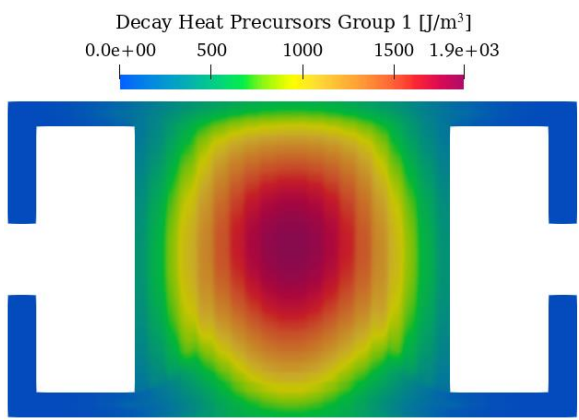

Figure 18: Time-averaged DHP concentration of group 1 calculated with NekRS.

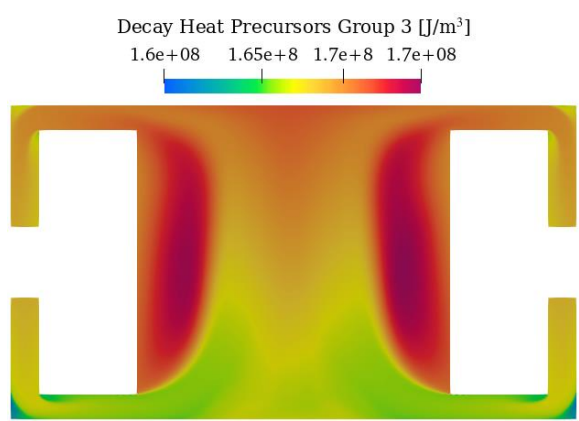

Figure 19: Time-averaged DHP concentration of group 3 calculated with NekRS.

## 8. CONCLUSIONS

A multiphysics model of the MSFR has been developed in Cardinal, considering neutronic – thermal hydraulics feedback and the presence of DNPs and DHPs. Neutronic equations were solved in OpenMC, while mass, momentum, energy, DNPs, and DHPs equations were solved in NekRS. Current limitations in OpenMC do not allow the modification of delayed neutron source locations or intensity, so the model has no DNP feedback. Preliminary results showed reasonable behavior for heat source, temperature, velocity, and DNP and DHP distributions. Low-velocity regions act as stagnation zones where the temperature reaches its maximum value.

For future work, we plan to include modifications in OpenMC to modify delayed neutron source locations. Additionally, we plan to improve the modeling presented here further by performing a sensitivity analysis and including additional effects that are expected to improve accuracy.